\documentclass[sigconf]{acmart}
\AtBeginDocument{%
  }

\setcopyright{acmlicensed}
\acmDOI{10.1145/3773966.3779397}
\acmISBN{978-1-4503-XXXX-X/18/06}

\usepackage{booktabs}
\usepackage{graphicx}
\usepackage{titlesec}
\usepackage{svg}
\titlespacing*{\section}{0pt}{4pt}{4pt} 
\titlespacing*{\subsection}{0pt}{3pt}{3pt} 
\usepackage{amsmath}
\usepackage{booktabs}
\usepackage{multirow} 
\usepackage{balance}
\usepackage{xcolor}
\usepackage{colortbl}  
\usepackage{textgreek}
\titlespacing*{\section}{0pt}{*1}{*1} 
\titlespacing*{\subsection}{0pt}{*1}{*1} 
\copyrightyear{2026}
\acmYear{2026}
\setcopyright{cc}
\setcctype{by}
\acmConference[WSDM '26]{Proceedings of the Nineteenth ACM International Conference on Web Search and Data Mining}{February 22--26, 2026}{Boise, ID, USA}
\acmBooktitle{Proceedings of the Nineteenth ACM International Conference on Web Search and Data Mining (WSDM '26), February 22--26, 2026, Boise, ID, USA}
\acmPrice{}
\acmDOI{10.1145/3773966.3779397}
\acmISBN{979-8-4007-2292-9/2026/02}

\begin{document}

\title{TRUE: A Reproducible Framework for LLM-Driven Relevance Judgment in Information Retrieval}

%
\author{Mouly Dewan}
\affiliation{%
  \institution{University of Washington}
  \city{Seattle, WA}
  \country{United States}}
\email{mdewan@uw.edu}

\author{Jiqun Liu}
\affiliation{%
  \institution{The University of Oklahoma}
  \city{Norman, OK}
  \country{United States}}
\email{jiqunliu@ou.edu}

\author{Chirag Shah}
\affiliation{%
  \institution{University of Washington}
  \city{Seattle, WA}
  \country{United States}}
\email{chirags@uw.edu}

\begin{abstract}
LLM-based relevance judgment generation has become a crucial approach in advancing evaluation methodologies in Information Retrieval (IR). It has progressed significantly, often showing high correlation with human judgments as reflected in LLMJudge leaderboards. However, existing methods for relevance judgments, rely heavily on sensitive prompting strategies, lacking standardized workflows for generating reliable labels. To fill this gap, we reintroduce our method, \textit{Task-aware Rubric-based Evaluation} (TRUE), for relevance judgment generation. Originally developed for usefulness evaluation in search sessions, we extend TRUE to mitigate the gap in relevance judgment due to its demonstrated effectiveness and reproducible workflow. This framework leverages iterative data sampling and reasoning to evaluate relevance judgments across multiple factors including intent, coverage, specificity, accuracy and usefulness. In this paper, we evaluate TRUE on the TREC DL 2019, 2020 and LLMJudge datasets and our results show that TRUE achieves strong performance on the system-ranking LLM leaderboards. The primary focus of this work is to provide a reproducible framework for LLM-based relevance judgments, and we further analyze the effectiveness of TRUE across multiple dimensions.
\end{abstract}

\ccsdesc[500]{Information systems~Language models}

\keywords{Information Retrieval, Usefulness Judgment, LLM Evaluation}

\maketitle

\begin{figure}[t]
    \centering
    \vspace{0pt}
    \includegraphics[width=1\linewidth]{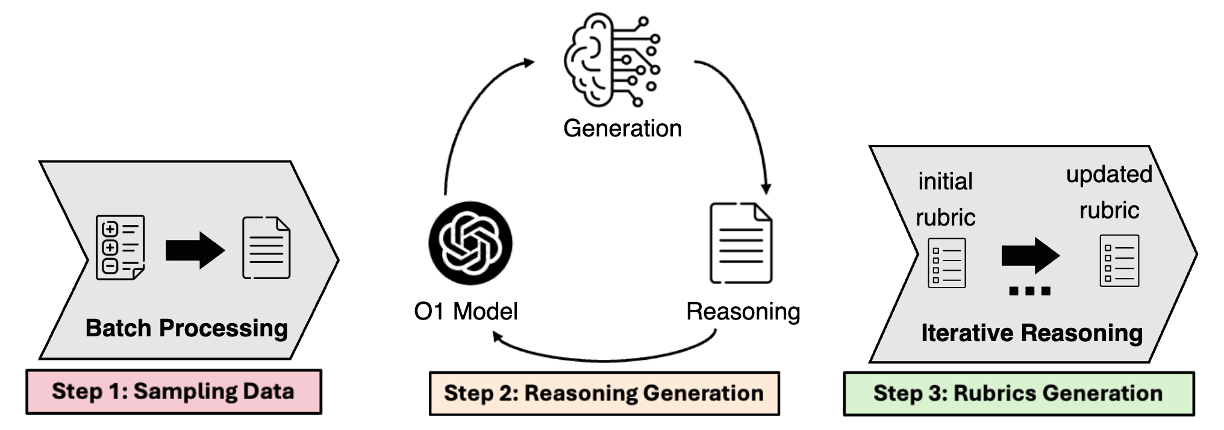}  
    \caption{Rubrics formulation with TRUE (Initial 3 steps).}
    \label{fig:true_formulation}
\end{figure}

\section{Introduction}

In the Information Retrieval (IR) domain, relevance assessment plays a crucial role in training and evaluating retrieval systems. Relevance labels are commonly generated by assessing a query together with its relevant passages following the Cranfield paradigm \cite{cleverdon1960aslib, cleverdon1967cranfield}. Such labels are typically assigned by human assessors, making the process costly and time-consuming. To address the limitations of traditional evaluation \cite{liu2022toward, jiang2025landscape}, IR researchers have increasingly explored automated LLM-based approaches to search evaluation \cite{thomas_large_2024, faggioli_perspectives_2023, chen_ai_2024, alaofi2024llms}. Recent advances show that LLMs can capture user intent and judge passage relevance with accuracy comparable to human assessors. Consequently, many approaches generate LLM-based relevance labels to construct system ranking leaderboards \cite{rahmani2025judging}. However, most existing methods rely on prompt-based procedures whose outcomes are highly sensitive to prompt wording and configuration. As a result, these methods often operate as black boxes with limited reproducibility \cite{farzi2025criteria}. Moreover, the lack of explicit rationales reduces transparency and makes relevance judgments difficult to verify and generalize across domains.

To enable consistent, reasoned, and reproducible evaluation, we introduce TRUE, a rubric-based framework for relevance assessment building on our prior work \cite{dewan2025llm}. Instead of directly judging query-passage pairs, TRUE evaluates documents using label-specific rubrics derived from key passage features. The framework operates in four steps: sampling the dataset, generating label-specific reasoning with OpenAI's o1 reasoning model, constructing rubrics for each relevance label, and applying these rubrics to assign relevance scores to passages. The rubrics capture five key features: intent alignment, coverage, specificity, accuracy and usefulness which make relevance judgments auditable and stable across prompts and datasets. We evaluate TRUE for LLM-based relevance judgment through the following research questions:

{\bfseries RQ1}: To what extent can TRUE accurately judge relevance compared to existing state-of-the-art methods?

{\bfseries RQ2}: To what extent does TRUE correspond to manual judgments?

{\bfseries RQ3}: How do the LLM models and retrieval evaluation metrics affect relevance judgment of TRUE?

\section{Related Works}

Recent work has increasingly explored the use of large language models (LLMs) for automated relevance assessment in information retrieval \cite{abbasiantaeb_can_2024, faggioli_perspectives_2023, rahmani2024llmjudge}. Thomas et al. \cite{thomas_large_2024} demonstrate that direct prompting enables LLMs to generate effective relevance judgments, establishing a baseline for automated evaluation frameworks such as UMBRELA \cite{upadhyay_large-scale_2024}. Arabzadeh and Clarke \cite{arabzadeh2025benchmarking} provide a comparative analysis of LLM-based methods including binary, graded, pairwise and nugget-based assessments on the TREC DL datasets. They report that pairwise preference judgments align most closely with human labels, while binary and graded methods such as UMBRELA, achieve stronger correlations with system rankings. Complementary work highlights the adaptability of LLMs: Sun et al. \cite{sun2023chatgpt} show their potential in zero-shot settings, while MacAvaney et al. \cite{macavaney2023one} explore one-shot learning for binary preference judgments. For reproducible judgment, Farzi et al. \cite{farzi2025criteria} introduce a multi-criteria relevance decomposition that addresses the limitations of direct grading approaches.  Recently, for LLM based assessments, user-centric usefulness judgment along with relevance have been explored as well \cite{dewan2025llm}. Both Faggioli et al. \cite{faggioli_perspectives_2023, abbasiantaeb_can_2024} find that LLM-based assessments are promising but caution against pitfalls such as mismatches between predicted labels and human-perceived relevance. In summary, the evidence shows that single, simplified metrics fall short of capturing the \textit{multi-faceted} aspects of relevance, which we further investigate in this paper.

\section{Methodology} 
Our main goal is to evaluate the extent to which LLMs can generate reproducible relevance labels that are highly similar to the human leaderboard system rankings. We elaborate our methods as:

\subsection{TRUE Formation}
We propose TRUE, a Task-aware Rubric based Evaluation method inspired by SAT/DSAT rubrics for user satisfaction in conversational AI systems by~\cite{lin2024interpretable}. Our method employs a four-phase prompting pipeline shown in Figure~\ref{fig:true_formulation}: data sampling, reasoning generation, rubric summarization and relevance generation with TRUE. This iterative design is intended to promote improvement, generalization and interpretability. The four key steps are explained below: 

\textbf{Step 1. Sampling Data:} We begin with batches of query document pairs from the datasets using 10\% of each dataset with equal distribution for each relevance label (0-3) and input them into OpenAI's o1 reasoning model (snapshot: o1-2024-12-17) \cite{zhong2024evaluation}. The model generates reasoning for relevance labels by identifying patterns between query and document across examples. We sample a subset to induce rubrics by learning label-specific patterns without processing the full dataset

\textbf{Step 2. Iterative Reasoning:} We feed the generated reasoning back into the OpenAI's o1 reasoning model, asking it to extract structured rules for each label. The model forms rubrics based on 5 key features for passage evaluation grouped into intent alignment, coverage, specificity, accuracy and usefulness. 

\textbf{Step 3. Rubric Formation:} After N = 3 iterations (\textit{N is adaptive}), to improve coverage and clarity, we finalize TRUE which includes reasoning rubrics for labels 0 to 3. During this process, we apply limited manual refinement to rephrase and standardize the rubrics for improved interpretability without changing their underlying semantic criteria. As mentioned previously, the rubrics classify rules into five key categories which helps evaluate the query-document pair for relevance. 

\textbf{Step 4. Relevance Generation:} We use the final rubric for each relevance label (0-3) categorized in five key features to generate relevance label. With the obtained relevance, we curate an LLMJudge leaderboard consistent with the most recent LLMJudge challenge. We additionally curate system rankings and compare our method TRUE with state-of-the-art relevance assessment methods.
The rubric evaluation is based on five dimensions: 1) Intent Alignment: how well the passage matches the user’s underlying query intent. 2) Coverage: how fully the passage addresses the information need (breadth and depth) \cite{saracevic2007relevance}. 3) Specificity: how precise, focused and contextually connected the passage is to the query \cite{saracevic2007relevance}. 4) Accuracy: whether the information is factually correct, reliable and not misleading and 5) Usefulness: whether the passage would actually help the user achieve their goal or stop searching \cite{belkin2008relevance}.

\begin{table}[t]
  \caption{Summary statistics for the Deep Learning Tracks 2019, 2020, LLMJudge and their label counts.}
  \label{tab:qrels-label-counts}
  \centering
\small
  \begin{tabular}{lrrrrrr}
    \toprule
    \textbf{qrels track} & \textbf{runs} & \textbf{topics} &
    \multicolumn{4}{c}{\textbf{Total counts for labels}} \\
    \cmidrule(lr){4-7}
     &  &  & \textbf{0} & \textbf{1} & \textbf{2} & \textbf{3} \\
    \midrule
    TREC DL 2019   & 36 & 43 & 5168 & 1601 & 1804 &  697 \\
    TREC DL 2020   & 59 & 54 & 7780 & 1940 & 1020 &  646 \\
    LLMJudge  & 35 & 25 & 4506 & 1403 &  625 &  697 \\
    \bottomrule
  \end{tabular}
\end{table}

\subsection{Datasets}
We conduct experiments on the test sets of three major TREC datasets (DL19, DL20 and LLMJudge) to ensure robust evaluation across varied time periods, different passage and query types. \textit{LLMJudge} \cite{rahmani2024llmjudge} comprises 25 topics and 35 runs (Table~\ref{tab:qrels-label-counts}), derived from system submissions to the TREC Deep Learning Track 2023 \cite{craswell2025overview}. We select this dataset as it is the most recent public test collection and reduces the risk of training-time data leakage for LLMs. For the LLMJudge dataset, we did not have a direct mapping to the TREC DL’23 passages. Therefore, we relied on the passage 'text' itself to establish the mapping, resulting in 6,327 query–passage pairs. \textit{ TREC DL20 \cite{DBLP:conf/trec/CraswellMMYC20}} has 54 queries, 59 systems for passage retrieval from TREC Deep Learning track of 2020.\textit{ TREC DL19 \cite{voorhees2020overview}} has 43 queries, 36 systems for passage retrieval from TREC Deep Learning track of 2019. The diverse datatsets along with its human assessor's label distribution has been shown in Table~\ref{tab:qrels-label-counts}.

\begin{table*}[h]
\centering
\small
\setlength{\tabcolsep}{7pt} 
\caption{Comparison of TRUE with established baselines and leaderboard correlation measured in Spearman’s rank (S) and Kendall’s Tau (T). Bold marks: (i) the top baseline scores, and (ii) our methods when they outperform both the best baseline and Multi-Criteria, to our knowledge the only framework that focuses on interpretable relevance generation.}
\label{tab:comparison}
\begin{tabular}{lcccccccccccc}
\toprule
 & \multicolumn{4}{c}{\textbf{DL2019}} & \multicolumn{4}{c}{\textbf{DL2020}} & \multicolumn{4}{c}{\textbf{LLMJudge}} \\
\midrule
 & \multicolumn{2}{c}{NDCG@10} & MAP & MRR & \multicolumn{2}{c}{NDCG@10} & MAP & MRR & \multicolumn{2}{c}{NDCG@10} & MAP & MRR \\
\cmidrule(lr){2-3} \cmidrule(lr){4-4} \cmidrule(lr){5-5}
\cmidrule(lr){6-7} \cmidrule(lr){8-8} \cmidrule(lr){9-9}
\cmidrule(lr){10-11} \cmidrule(lr){12-12} \cmidrule(lr){13-13}
 & S & T & S & S & S & T & S & S & S & T & S & S \\
\midrule
\textbf{LLaMA 3.3 70B} & & & & & & & & & & & & \\
\quad TRUE (Ours)        & \textbf{0.965} & \textbf{0.857} & \textbf{0.762} & 0.822 & \textbf{0.988} & 0.919 & 0.783 & 0.725 & 0.961 & 0.845 & 0.916 & 0.343 \\
\quad TRUE Reasoning (Ours)        & \textbf{0.969} & \textbf{0.863} & 0.749 & 0.856 & \textbf{0.988} & 0.919 & 0.687 & 0.704 & 0.960 & 0.849 & 0.910 & 0.351 \\
\quad Multi-Criteria~\cite{farzi2025criteria} & 0.967 & 0.856 & 0.762 & 0.881 & \textbf{0.988} & 0.927 & \textbf{0.919} & 0.921 & \textbf{0.994} & \textbf{0.952} & 0.968 & 0.911 \\

\quad Thomas~\cite{thomas_large_2024}        & 0.969 & 0.863 & \textbf{0.873} & \textbf{0.925} & 0.922 & \textbf{0.929} & 0.821 & 0.887 & 0.991 & 0.939 & \textbf{0.970} & 0.902 \\
\quad Faggioli~\cite{faggioli_perspectives_2023} & 0.927 & 0.767 & 0.703 & 0.857 & 0.980 & 0.901 & 0.857 & \textbf{0.963} & 0.985 & 0.912 & 0.968 & \textbf{0.958} \\
\quad UMBRELA~\cite{10.1145/3726302.3730317, upadhyay_umbrela_2024}   & \textbf{0.970} & \textbf{0.869} & -- & -- & 0.986 & 0.911 & -- & -- & 0.993 & 0.946 & -- & -- \\
\midrule
\textbf{LLaMA 3 8B} & & & & & & & & & & & & \\
\quad TRUE (Ours)        & \textbf{0.986} & \textbf{0.921} & \textbf{0.627} & 0.623 & 0.960 & 0.840 & 0.589 & 0.567 & 0.730 & 0.556 & 0.890 & 0.367 \\
\quad TRUE Reasoning (Ours)        & \textbf{0.979} & \textbf{0.894} & \textbf{0.696} & 0.650 & \textbf{0.976} & \textbf{0.873} & \textbf{0.638} & 0.611 & 0.876 & 0.701 & \textbf{0.910} & 0.300 \\
\quad Multi-Criteria~\cite{farzi2025criteria} & 0.972 & 0.871 & 0.349 & 0.852 & 0.972 & 0.871 & 0.617 & 0.626 & \textbf{0.994} & \textbf{0.952} & \textbf{0.968} & 0.874 \\

\quad Thomas~\cite{thomas_large_2024}        & 0.971 & 0.879 & 0.549 & 0.898 & 0.973 & 0.870 & \textbf{0.917} & \textbf{0.891} & 0.987 & 0.915 & 0.964 & 0.833 \\
\quad Faggioli~\cite{faggioli_perspectives_2023} & 0.974 & 0.874 & \textbf{0.750} & \textbf{0.909} & \textbf{0.982} & \textbf{0.909} & 0.902 & 0.866 & 0.975 & 0.892 & 0.963 & \textbf{0.943} \\
\quad UMBRELA~\cite{10.1145/3726302.3730317, upadhyay_umbrela_2024}   & \textbf{0.975} & \textbf{0.894} & -- & -- & 0.973 & 0.870 & -- & -- & 0.989 & 0.931 & -- & -- \\
\bottomrule
\end{tabular}
\label{tab:correlations}
\end{table*}

\subsection{Experimental Setup}
Our evaluation uses two open LLMs from the same family: LLaMA 3 8B and LLaMA 3.3 70B \cite{dubey2024llama}. We include LLaMA 3.3 70B as a recent large-scale public model and LLaMA 3 8B as a medium-scale comparison point. We employ two prompting strategies for relevance generation. (i) \textbf{TRUE} applies the rubric directly to produce a relevance label. (ii) \textbf{TRUE Reasoning} elicits reasoning similar to chain-of-thought (CoT) \cite{wei2022chain} across the five dimensions of the rubric and then determines a final score. All information was passed using zero-shot prompting, inspired by the DNA prompt for relevance assessment \cite{thomas_large_2024}\footnotemark[1]\footnotetext[1]{Prompts: https://github.com/moulydewan/TRUE-Relevance-Judgment.git}. We ask the LLM to label the relevance of documents given the respective query as: 3 - Highly Relevant, 2 - Fairly Relevant, 1 - Partially Relevant, and 0 - Not Relevant at all. For model evaluation, we experiment with LLaMA 3 8B and LLaMA 3.3 70B through the API service from AWS Bedrock. We follow model parameter settings from \cite{thomas_large_2024} with temperature set to 0 and top\_p to 1.

\section{Results}
We analyze and evaluate TRUE along leaderboard correlation, system ranking performance and comparison with manual judgment. In this section, we discuss the research questions in depth.
\subsection{RQ1: Comparison of TRUE baselines}
We compare TRUE with established baselines such as Multi-Criteria \cite{farzi2025criteria}, Thomas \cite{thomas_large_2024}, Faggioli \cite{faggioli_perspectives_2023} and UMBRELA \cite{10.1145/3726302.3730317, upadhyay_umbrela_2024}. Among these methods, all of them approach relevance judgment through direct zero-shot prompting. Except Multi-Criteria is one of the highlighted method that focuses on more robust evaluation approach considering features for relevance judgment. The difference between our method and Multi-criteria is that we focus more towards reproducibility by sampling and generating reasoning before the formation of score rubrics. Our entire framework focuses on chain-of-thought (CoT) reasoning before rubric formation and also while generating the final relevance score. TRUE also expands its key features to intent alignment, coverage, specificity, accuracy and usefulness covering all possible dimensions of relevance judgment \cite{saracevic2007relevance}. We evaluate the performance of TRUE across three datasets for two LLM models across trec\_eval metrics like NDCG@10, mean average precision (MAP) and mean reciprocal rank (MRR) shown in Table~\ref{tab:comparison}. We use correlation metrics Spearman's $\rho$ and Kendall’s $\tau$ to compare the performance. In DL19, for LLaMA 3.3 70B both TRUE and TRUE Reasoning achieve very high leaderboard correlations (NDCG@10), matching or slightly below UMBRELA. On MAP and MRR, they are competitive with Multi-Criteria, but Thomas achieves the strongest MAP/MRR. For Llama 3 8B, both TRUE methods outperform all baselines in NDCG@10 correlation. For MAP, both perform better than Mutli-Criteria and slightly improves over TRUE (0.627/0.623), but remain lower than the best baseline. For DL20, in LLama 3.3 70B, both TRUE achieve best NDCG@10 correlations (0.988). MAP and MRR are lower, trailing behind Multi-Criteria. In Llama 3 8B, TRUE Reasoning (0.976/0.873) again show strong leaderboard alignment more than Mutli-Criteria but underperforms in MAP/MRR. On the newest benchmark (LLMJudge), TRUE methods remain competitive in leaderboard correlation but do not surpass strong baselines in MAP/MRR. Overall, for Llama 3 8B, TRUE-Reasoning improves correlation robustness (DL2019/2020) but the reasoning step does not consistently translate to higher MAP/MRR. All reported correlations are statistically significant ($p < 0.05$).


\begin{figure*}[t]
    \centering
    \includegraphics[width=0.7\linewidth]{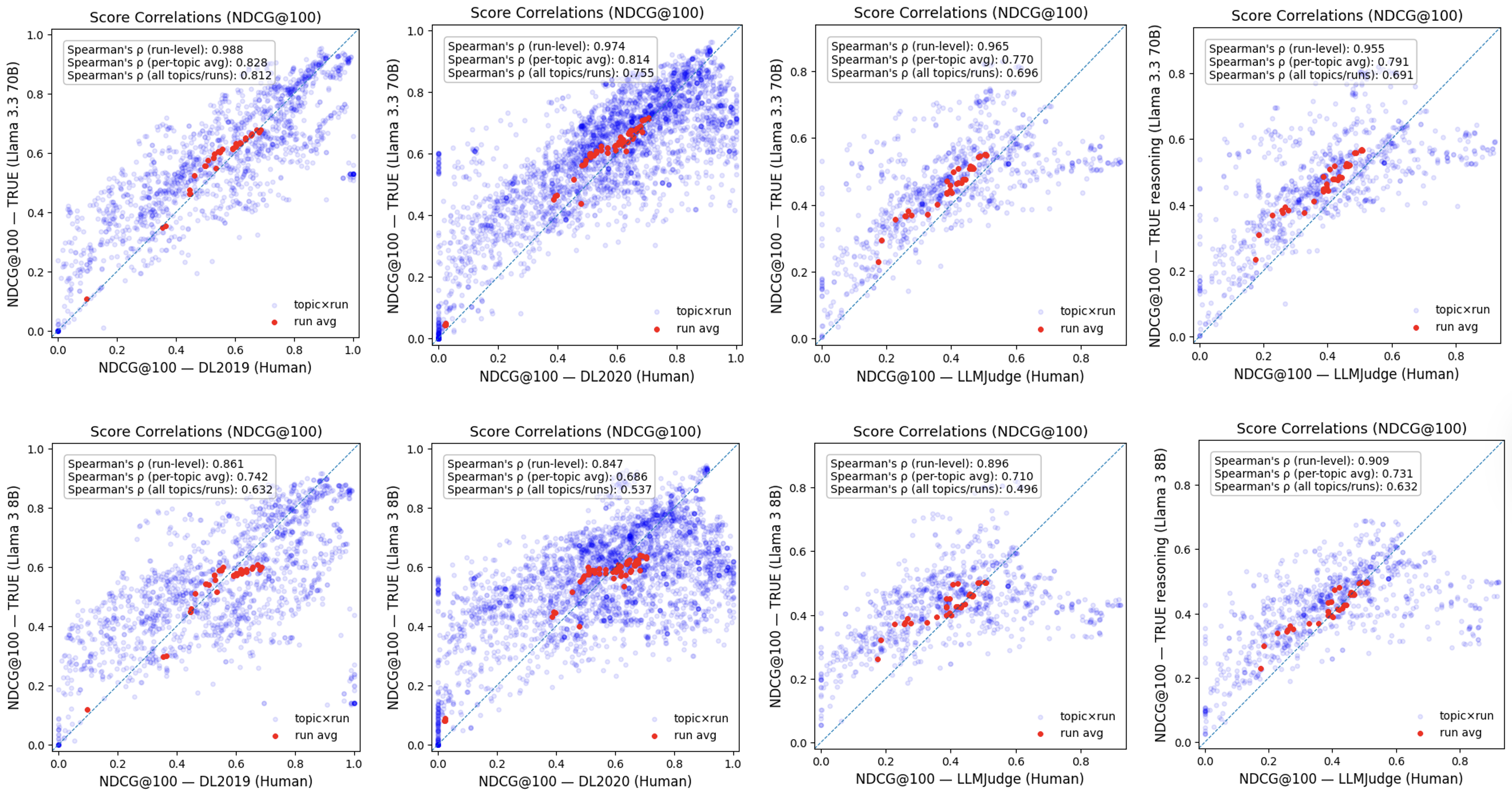}
    \Description{Graph comparing TRUE across LLMs and between TRUE and TRUE Reasoning on LLMJudge.}
    \caption{Comparisons of TRUE across LLMs (first 3 columns) and between TRUE and TRUE Reasoning on LLMJudge (last 2 columns: Llama 3.3 70B). Red = run-level scores and blue = topic/run pairs with rank correlations in terms of Spearman’s ($\rho$).}
    \label{fig:graph2}
\end{figure*}


\begin{table}[h]
\centering
\small
\caption{Cohen's ($\kappa$) scores for TRUE on usefulness and relevance datasets. Top 3 scores in each column is marked bold.}
\begin{tabular}{llcccc}
\toprule
\textbf{Domain} & \textbf{Dataset} & \textbf{4-point} & \textbf{0\textbar123} & \textbf{01\textbar23} & \textbf{012\textbar3} \\
\midrule
Usefulness \cite{dewan2025llm} & THUIR \cite{mao_understanding_2017}     & \textbf{0.262} & \textbf{0.352} & \textbf{0.434} & \textbf{0.389} \\
Usefulness \cite{dewan2025llm} & QRef \cite{chen2021towards}       & 0.119 & 0.079 & 0.245 & 0.220 \\
Relevance  & DL2019     & \textbf{0.251} & \textbf{0.382} & \textbf{0.456} & \textbf{0.274} \\
Relevance  & DL2020     & \textbf{0.219} & \textbf{0.350} & 0.350 & \textbf{0.315} \\
Relevance  & LLM4Eval   & 0.197 & 0.321 & \textbf{0.386} & 0.238 \\
\bottomrule
\end{tabular}
\label{tab:agreement}
\end{table}

\subsection{RQ2: TRUE and human judgment}
In this section, we examine fine-grained agreement between human judgments and LLMs for relevance. In Table~\ref{tab:agreement}, for consistency with LLMJudge challenge \cite{rahmani_llm4eval_2024} and state-of-the-art methods, we report inter-rater agreement using Cohen’s $\kappa$ between ground-truth human labels and TRUE. We also evaluate TRUE on usefulness judgment on our earliest work \cite{dewan2025llm}. Evaluating TRUE on usefulness is challenging due to the lack of standardized datasets and metrics. To address this, we also report results on two usefulness datasets (THUIR, QREF). Overall agreement levels are moderate, Cohen's $\kappa \approx (0.20 - 0.45$) consistent with ranges reported in the LLMJudge challenge. Agreement is higher when labels are collapsed into broader categories (0|123, 01|23). THUIR shows the highest consistency, with Cohen's $\kappa$ up to 0.434 (01|23) split but QRef scores are lower suggesting noisier labeling. Similar interpretations has been carried out in the original usefulness judgment as well verifying its effectiveness. DL19 yields the strongest agreement among relevance, Cohen's $\kappa$ up to 0.456 for (01|23) split. It aligns best on DL19 (relevance) and THUIR (usefulness), particularly when judgments are collapsed into binary high vs. low (01|23) split for relevance and usefulness categories.

\subsection{RQ3: Metrics and models in TRUE} We evaluate the effect of different LLMs along NDCG@100 focusing on  Spearman's $\rho$. In Figure~\ref{fig:graph2}, we show the agreement on a scatter plot where the center diagonal line means perfect agreement between human vs LLMs. The blue dots represent each topic per run and the red dots represent the average run of the topics. First three columns showcases the Spearman's rank across DL19, DL20 and LLMJudge. In top row representing LLama 3.3 70B, for DL19, DL20 and LLMJudge, the Spearman’s $\rho$ values are high (0.988, 0.974, 0.965), indicating strong monotonic correlation between TRUE and human judgments. From the distribution of red dots we can see that its identical to the center diagonal line for all. For LLaMA 3 8B in bottom row correlations are noticeably lower, particularly on DL20 (0.847). This suggests that the smaller model struggles more to match human system-level ranking. Across all datasets, run-level correlations are high but topic-level correlations are consistently lower. This indicates that while TRUE captures system-level ranking very well, fine-grained agreement at the topic/query level varies. For last two columns we choose LLMJudge for TRUE vs. TRUE Reasoning. For LLaMa 3 8B, reasoning improves performance which can be noticeable in the plot with red dots smoothening out along the center line. Therefore, reasoning has limited impact for large models but noticeably helps smaller models, closing some of the performance gap. These results indicate that TRUE is robust for system-level evaluation where rank correlation with human ground truth matters most.


\section{Discussion and Conclusion}

In this paper, we investigate TRUE for automated relevance judgments and present a reproducible, chain-of-thought (CoT) framework. We evaluate TRUE against established baselines, examine the impact of model families and report standard trec\_eval metrics such as NDCG@10, NDCG@100, MAP, and MRR. We also assess inter-rater agreement between TRUE and human judgments. Across these analyses, TRUE delivers robust relevance assessments, in line with existing methods. Several challenges in designing TRUE were addressed using the LLM evaluation trope framework \cite{dietz2025principles}. To mitigate \textit{circularity}~\cite{rahmani2025judging}, we employ two different LLMs, for reasoning generation and evaluation. This also reduces \textit{memorization} by exposing only a sample of the data to derive rubrics. In our framework, reasoning is generated with OpenAI's o1 reasoning model, while evaluation is performed by models from the LLaMA family. While comparison of TRUE with other baselines gave positive results only for certain metrics, MAP and MRR are comparatively low due to noisy label distribution. LLM-generated labels exhibit a central tendency bias, concentrating scores and depressing rank-sensitive metrics. This framework advances reproducible assessment within the \textit{LLMs-as-Judges} domain~\cite{gu2024survey} by systematically probing how LLMs judge relevance and analyzing both their strengths and limitations. Building on these insights, for future work we plan to pursue task-specific and session aware relevance judgment~\cite[e.g.][]{markwald_constructing_2023, liu2020identifying} including fine-tuning, as well as the development of specialized evaluators for more accurate relevance and robust labeling~\cite{markwald_constructing_2023, wang2024task}. In addition, we will extend TRUE to automated relevance judgments in text and conversational retrieval systems, enabling real-time \textit{relevance and usefulness based} re-ranking, and design systems that recommend effective and interpretable search paths~\cite{liu2022leveraging, yu2025chat}.

\bibliographystyle{ACM-Reference-Format}
\balance
\bibliography{reference}

\end{document}